\documentclass[twocolumn]{aastex63}

\received{}
\revised{}
\accepted{}
\submitjournal{PASP}

\shorttitle{Ni-masses in SNe Ia}
\shortauthors{Bora, Vink\'o \& K\"onyves-T\'oth}

\begin{document}
\title{Initial $^{56}$Ni Masses in Type Ia Supernovae}
\correspondingauthor{J. Vink\'o}
\email{vinko@konkoly.hu}

\author[0000-0001-6232-9352]{Zs\'ofia Bora}
\affiliation{ELTE E\"otv\"os Lor\'and University, Department of Astronomy, P\'azm\'any P\'eter s\'et\'any 1/A, Budapest, 1117 Hungary}

\author[0000-0001-8764-7832]{J\'ozsef Vink\'o}
\affiliation{ Konkoly Observatory,  CSFK, Konkoly-Thege M. \'ut 15-17, 
Budapest, 1121, Hungary}
\affiliation{ELTE E\"otv\"os Lor\'and University, Institute of Physics, P\'azm\'any P\'eter s\'et\'any 1/A, Budapest, 1117 Hungary}
\affiliation{Department of Optics \& Quantum Electronics, University of Szeged, D\'om t\'er 9, Szeged, 6720, Hungary}

\author[0000-0002-8770-6764]{R\'eka K\"onyves-T\'oth}
\affiliation{ Konkoly Observatory,  CSFK, Konkoly-Thege M. \'ut 15-17, 
Budapest, 1121, Hungary}

\begin{abstract}

We infer initial masses of the synthesized radioactive nickel-56 in a sample of recent Type Ia supernovae applying a new formalism introduced recently by Khatami \& Kasen (2019). It is shown that the nickel masses we derive do not differ significantly from previous estimates based on the traditional Arnett-model. We derive the $\beta$ parameter for our sample SNe and show that these are consistent with the fiducial value of $\sim 1.6$ given by Khatami \& Kasen (2019) from SN Ia hydrodynamical simulations. 

\end{abstract}


\section{Introduction} \label{sec:intro}

Type Ia supernovae (SNe Ia) are among the most suitable objects for extragalactic distance measurements \citep[see e.g.][for a recent review]{jha19}. Although the well-known correlation between their peak absolute brightness (e.g. $M_B$ in the $B$-band) and the light curve decline rate (e.g. $\Delta m_{15}$) is still based on empirical calibrations, it was realized decades ago that the peak brightness is connected to the amount of radioactive nickel ($^{56}$Ni) synthesized during the thermonuclear explosion of the carbon-oxygen white dwarf {(C/O WD; \citealt{arnett82, gold18})}. Understanding the physics behind the peak brightness - decline rate - nickel mass ($M_{\rm Ni}$) connection may have a crucial importance in improving the distance measurement methods. 

The nickel yield of SNe Ia may also be the smoking gun for the explosion mechanism triggering the thermonuclear runaway in the C/O white dwarf \citep[see e.g.][and references therein]{seit17}. For example, delayed-detonation models that assume the explosion of near-Chandrasekhar mass ($M_{Ch}$) WDs predict higher $^{57}$Ni/$^{56}$Ni abundance ratio than double-detonation or violent merger models in which the exploding WD has sub-$M_{Ch}$ mass. \citet{graur16} found about twice of the solar value for the $^{57}$Ni/$^{56}$Ni mass ratio from the late-time light curve decline rate of SN~2012cg, which led to the conclusion that the exploding star in SN~2012cg was a near-$M_{Ch}$ WD. On the other hand, \citet{scalzo14b} pointed out that the observed diversity of both the $^{56}$Ni- and the ejecta masses of SNe Ia ($0.3 < M_{\rm Ni} < 0.8$ M$_\odot$, and $0.8 < M_{\rm ej} < 1.5$ M$_\odot$, respectively) suggests that sub-$M_{Ch}$ explosion mechanisms, such as the double detonation or the violent merger scenario  are responsible for the bulk of the observed SNe Ia \citep[see e.g.][for a review of explosion models]{maoz14}. Similar conclusions have been presented by numerous other studies, e.g. in \citet{childress15}, \citet{dhawan18}, \citet{scalzo19}, \citet{wygoda19} and \citet{konyvestoth20}. 

The estimation of the initial amount of $M_{\rm Ni}$ synthesized during the explosion is traditionally based on Arnett's rule \citep{arnett82}, which states that at the moment of maximum luminosity the instantaneous energy input from the radioactive $^{56}$Ni $\rightarrow$ $^{56}$Co $\rightarrow$ $^{56}$Fe decay chain is equal to the peak bolometric luminosity. Since the energy generation rate from radioactive Ni-decay depends linearly on the initial amount $^{56}$Ni, it provides a relatively easy way of inferring $M_{\rm Ni}$ \citep[e.g.]{stritz06}. Modeling the unblended Fe and Co features in the nebular spectra of SNe Ia is an alternative method of deriving the Ni-mass \citep[e.g.][]{mazzali97, childress15}. \citet{stritz06} showed that these two methods, i.e. Arnett's rule and the nebular spectral modeling, provide consistent Ni-mass estimates. 

Recently, \citet{khatami19} found that Arnett's rule has limited accuracy. The main reason for this is that in the classical Arnett-model the energy density profile of the ejecta is assumed to be self-similar immediately after explosion. In reality, however, if the heating source ($^{56}$Ni) is central, like in core-collapse SNe, then some time is needed to reach self-similarity. For more evenly mixed sources (like SNe Ia), self-similarity is reached earlier, so Arnett's rule applies to these objects quite well. \citet{khatami19} found that for a central heating source, Arnett's rule underestimates the peak luminosity, while for more evenly mixed heating it may give an overestimate. They also presented a new relation between the peak luminosity and peak time of SN light curves that does not assume self-similarity.

In this paper we compare the nickel mass estimates from the formulae of \citet{khatami19} to ones derived from recent observations of SNe Ia based on Arnett's rule \citep{konyvestoth20}. In Section~\ref{sec:method} we review the methodology of getting nickel mass estimates from the observations, then the results from the two methods are compared to each other in {\bf Section~\ref{sec:comparison}}. In Section~\ref{sec:results} we calibrate the value of $\beta$ using two other methods for determining the nickel mass of Type Ia SNe, which are independent from Arnett's rule. Finally, Section~\ref{sec:conclusion} summarizes our conclusions.   

\section{Methods}\label{sec:method}

In the original, self-similar model of a SN Ia, assuming homologous expansion of a constant density ejecta, Arnett's rule takes the following form: 
\begin{equation}
    \label{eq:1}
    L_{\rm peak} = \alpha L_{\rm heat}(t_{\rm peak}),
\end{equation}
where $L_{\rm heat}(t_{\rm peak})$ is the heating function at the moment of the luminosity peak, and $\alpha \sim 1$ is a correction factor that accounts for small deviations from the assumptions of the Arnett-model.

For SNe Ia $L_{heat}$ is assumed to be the usual exponential form of the decay chain of $^{56}$Ni $\rightarrow$ $^{56}$Co $\rightarrow$ $^{56}$Fe. {\bf If $\varepsilon_{ \rm Ni} = 7.9 \times 10^{43}$ erg~s$^{-1}$ and $\varepsilon_{ \rm Co} = 1.45 \times 10^{43}$ erg~s$^{-1}$ are the heating rates and $t_{ \rm Ni} = 8.8$ days and $t_{ \rm Co} = 111.3$ days are the {\bf e-folding} timescales of $^{56}$Ni and $^{56}$Co \citep[e.g.][]{bw17},}
then the heating function takes the form of
\begin{equation}
    \label{eq:2}
    L_{ \rm heat}(t) = \frac{M_{ \rm Ni}}{\rm M_\odot} \cdot \left[ (\varepsilon_{ \rm Ni} - \varepsilon_{ \rm Co})e^{-t/t_{ \rm Ni}} + \varepsilon_{ \rm Co} e^{-t/t_{ \rm Co}} \right].
\end{equation}

Since the heating function depends linearly on $M_{\rm Ni}$, it can be expressed simply as
\begin{equation}
\label{eq:3}
    M_{\rm Ni} ~=~ 
    \frac{L_{\rm peak}}
    {\alpha Q(t_{\rm peak})},
\end{equation}
where $Q(t)$ is the time-dependent part of the $L_{\rm heat}(t)$ function in Eq.\ref{eq:2}.

The caveat of this simple approach is that the inferred $M_{\rm Ni}$ depends critically on $t_{\rm peak}$ and $L_{\rm peak}$, which must be known accurately to get a reasonable result. If the photometric sampling is sparse, it may be difficult to get a precise $M_{\rm Ni}$. To overcome this difficulty, one possibility is to fit the entire bolometric light curve with the prediction of the Arnett-model \citep[e.g.][]{valenti08, manos12}. In this case the theoretical luminosity can be expressed as
\begin{equation}\label{eq:4}
    L(t) = \frac{2}{t_{\rm m}^2} 
    (1 - e^{-(t_{\gamma}/t)^2}) \int_0^t t' L_{\rm heat}(t') e^{(t'-t)^2/t_{\rm m}^2} dt' ,
\end{equation}
where $t_{\rm m}$ is the mean light curve timescale (close to, but not equal with $t_{\rm peak}$) and $t_{\gamma}$ is the timescale for the gamma-ray leaking \citep{arnett82, valenti08, manos12}. Inserting Equation~\ref{eq:2} into Equation~\ref{eq:4} and fitting it to the bolometric light curve around the peak, one can get the best-fit estimate for $M_{\rm Ni}$.   

\citet{khatami19} inferred a new relation between the peak luminosity of a SN and the spatial and temporal distribution of its radioactive heating source: 
\begin{equation}
    L_{ \rm peak} = \frac{2}{\beta^2 t_{ \rm peak}^2} \int_0^{\beta t_{ \rm peak}} t' L_{ \rm heat }(t') dt' ,
    \label{eq:5}
\end{equation}
where $\beta$ is a numerical parameter that is related to the spatial distribution of heating, while $L_{ \rm heat}(t)$, again, describes the type of heating that powers the SN ejecta. It can have many forms depending on the SN type \citep[see][for a list of different sources]{khatami19}. 

Inserting $L_{heat}(t)$ from Equation~\ref{eq:2} into Equation~\ref{eq:5}, 
$L_{ \rm peak}$ 
can be expressed as\footnote{In the original publication of \citet{khatami19} this formula was written with $1- \frac{\beta t_{ \rm peak}}{t_{ \rm Ni}} $ instead of $1- \left(\frac{\beta t_{ \rm peak}}{t_{ \rm Ni}} + 1\right) $ in the first part of the sum within the square brackets. Later it has been corrected in a new version uploaded to arxiv.org. We present and use the correct version here.}

\begin{eqnarray}
    L_{ \rm peak} & = \frac{ 2 M_{\rm Ni} \varepsilon_{\rm Ni} t_{\rm Ni}^2}{\beta^2 t_{\rm peak}^2} \cdot \left[ 
\left(1 - \frac{\varepsilon_{\rm Co}}{\varepsilon_{\rm Ni}} \cdot \frac{t_{\rm Co}}{t_{\rm Co} - t_{\rm Ni}} \right ) \right. \cdot \nonumber \\
    & \left( 1 - \left( \frac{\beta t_{\rm peak}}{t_{\rm Ni}} + 1 \right) e^{-\frac{\beta t_{\rm peak}}{t_{\rm Ni}}} \right) + \frac{t_{\rm Co}^2}{t_{\rm Ni}^2} \frac{\varepsilon_{\rm Co}}{\varepsilon_{\rm Ni}} \cdot \nonumber \\ 
    & \left. \frac{t_{\rm Co}}{t_{\rm Co} - t_{\rm Ni}} \cdot \left( 1 - \left(\frac{\beta t_{\rm peak}}{t_{\rm Co}} + 1 \right) e^{-\frac{\beta t_{\rm peak}}{t_{\rm Co}}}\right) \right] 
    \label{eq:6}
\end{eqnarray}   

\begin{figure}
    \centering
    \includegraphics[width=8cm]{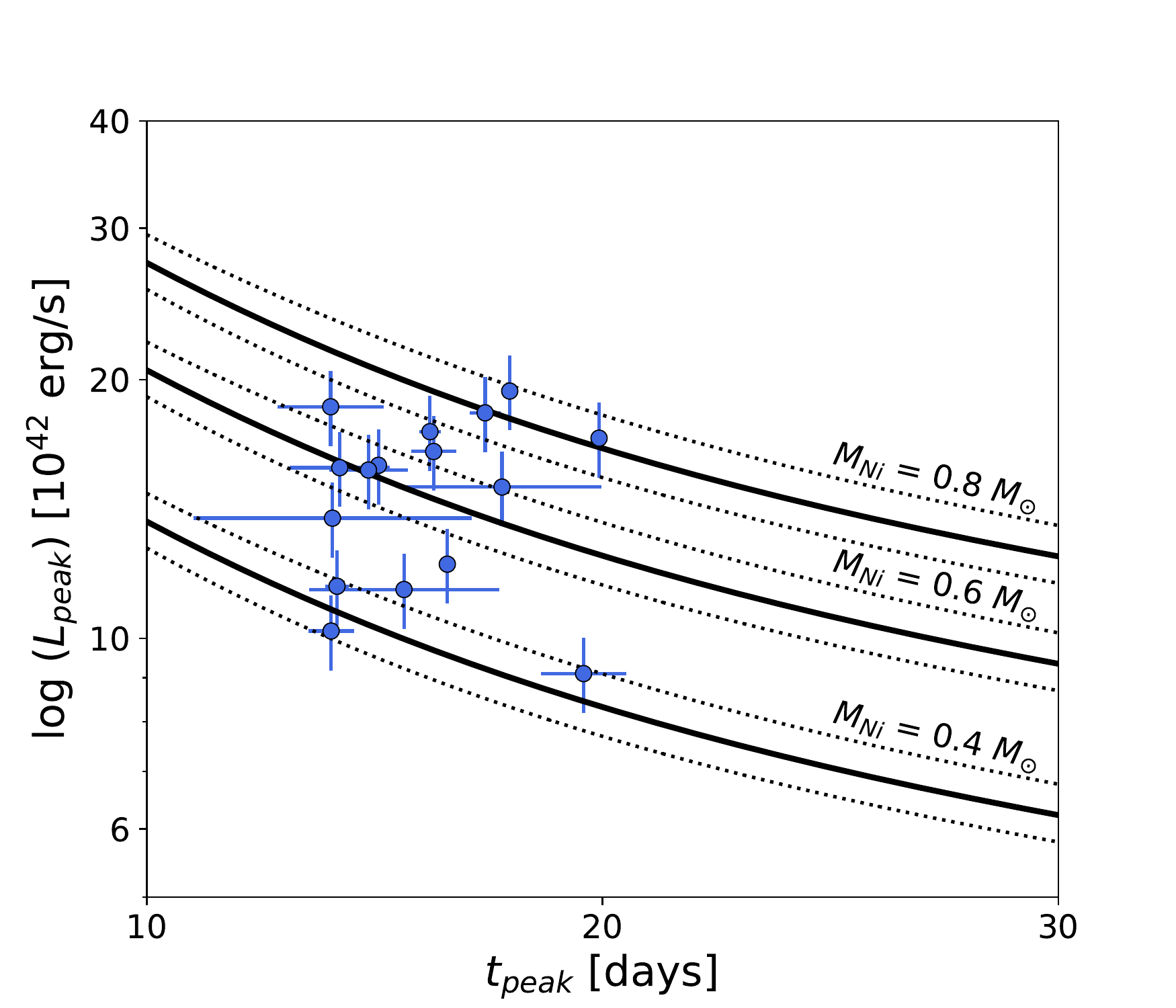}
     \caption{$L_{\rm peak}$ as a function of $t_{\rm peak}$ (Eq.~\ref{eq:6}) for $\beta$ = 1.666 $\pm$ 0.188 (as derived in Section~\ref{sec:results}) for different values of $M_{Ni}$. $L_{peak}$ is marked with solid lines, the dotted lines are showing the uncertainty caused by the standard deviation of $\beta$. The $L_{peak, obs}$ values from Table~\ref{tab:1} are shown as well. }
    \label{fig:1}
\end{figure}

Equation~\ref{eq:6} describes the dependence of the peak luminosity on $t_{\rm peak}$, i.e. the rise time to the bolometric maximum light, if the SN is powered by the radioactive decay of $^{56}$Ni and $^{56}$Co.
Again, the peak luminosity is linearly proportional to $M_{\rm Ni}$, but the connection between $L_{\rm peak}$ and $t_{\rm peak}$ is not as simple as in Equation~\ref{eq:1}. 
Figure~\ref{fig:1} shows the $L_{ \rm peak}$ vs. $t_{ \rm peak}$ relation {\bf for $\beta = 1.666 \pm 0.188$ (solid and dotted lines) assuming different values of $M_{\rm Ni}$, as well as the observed data given in Table~\ref{tab:1}.} 

\section{Comparison with observations}
\label{sec:comparison}

In this paper we use the peak time - luminosity relation expressed in Equation~\ref{eq:6} to infer the initial $^{56}$Ni mass of 16 SNe Ia studied recently by \cite{konyvestoth20}.  They estimated ejecta masses, initial nickel masses and other parameters by fitting bolometric light curves of 16 SNe Ia with the Arnett-model, as described in Section~\ref{sec:method}. 

\begin{table}
\centering
\begin{tabular}{cccc} 
\hline
Object & $t_{ \rm peak}$  &  $L_{ \rm peak, obs}$  & $t_{\gamma}$ \\ 
 & [days] & [$10^{43}$ erg/s]  &   [days]\\
\hline
 SN~2011fe   & 16.59 $\pm$0.06  & 1.22 $\pm$0.12 & 37.603 $\pm$0.670 \\         
 Gaia16alq  & 19.92 $\pm$0.42 & 1.71 $\pm$0.17 & 46.669  $\pm$0.717\\ 
 SN~2016asf  & 15.08 $\pm$2.58 & 1.59 $\pm$0.16 & 39.192 $\pm$1.329\\          
 SN~2016bln  & 17.42 $\pm$0.34 & 1.83 $\pm$0.18 & 44.508 $\pm$1.125\\         
 SN~2016coj  & 14.17 $\pm$0.26  & 1.15 $\pm$0.11 & 32.967 $\pm$0.863\\           
 SN~2016eoa  & 14.07 $\pm$3.05  & 1.38 $\pm$0.14 & 39.038 $\pm$0.935\\         
 SN~2016ffh  & 14.03 $\pm$1.16  & 1.86 $\pm$0.19 & 40.521 $\pm$0.926\\      
 SN~2016gcl  & 17.79 $\pm$2.18  & 1.50 $\pm$0.15 & 43.623 $\pm$1.182\\          
 SN~2016ixb  & 15.64 $\pm$2.09  & 1.14 $\pm$0.11 & 30.520 $\pm$1.345 \\           
 SN~2017cts  & 14.23 $\pm$1.10  & 1.58 $\pm$0.16 & 42.485 $\pm$0.959\\           
 SN~2017erp  & 17.96 $\pm$0.09  & 1.94 $\pm$0.19 & 37.603 $\pm$1.084\\            
 SN~2017fgc  & 16.21 $\pm$0.24  & 1.74 $\pm$0.17 & 45.398 $\pm$0.941\\            
 SN~2017fms  & 14.04 $\pm$0.50  & 1.02 $\pm$0.10 & 34.612 $\pm$0.731\\           
 SN~2017hjy  & 16.29 $\pm$0.49  & 1.65 $\pm$0.16 & 39.484 $\pm$0.840\\ 
 SN~2017igf  & 19.58 $\pm$0.93 & 0.91 $\pm$0.09  & 34.554 $\pm$1.193\\
 SN~2018oh   & 14.86 $\pm$0.86 & 1.57 $\pm$0.16 & 44.654 $\pm$0.928\\          
\hline
\end{tabular}
\caption{Parameters of the sample SNe Ia, adopted from \citet{konyvestoth20}.} 
\label{tab:1}
\end{table}

Table \ref{tab:1} summarizes the parameters for the sample SNe collected from \citet{konyvestoth20}: here $t_{ \rm peak}$ is the rise time from explosion to maximum luminosity in days, $L_{\rm peak,obs}$ is the peak luminosity determined from the observations.

Applying Equation~\ref{eq:6} to these data, we derived new $M_{Ni}$ values using the $t_{ \rm peak}$ and $L_{ \rm peak,obs}$ values from Table~\ref{tab:1}.  We adopted $\beta = 1.6$ as recommended by \citet{khatami19} for SNe Ia based on radiation hydrodynamical simulations. 

The results are shown in Table \ref{tab:2} as $M_{Ni}^{KK}$. Uncertainties are derived via propagating the errors listed in Table~\ref{tab:1}. 

\begin{table*}
\centering
\begin{tabular}{cccccc} 
\hline
  Object &   $M_{ \rm Ni}^{\rm Arnett}$  &   $M_{ \rm Ni}^{\rm tail}$ &   $M_{ \rm Ni}^{\rm t15}$ & $M_{ \rm Ni}^{\rm avg}$ &   $M_{ \rm Ni}^{\rm KK}$\\ 
   & $[M_{ \rm \odot}]$ & $[M_{ \rm \odot}]$ & $[M_{ \rm \odot}]$ & $[M_{ \rm \odot}]$  & $[M_{ \rm \odot}]$\\
\hline
SN~2011fe  &  0.567  $\pm$0.042  & 0.581  $\pm$0.049 &  0.59  $\pm$0.051  &  0.579  $\pm$0.047  &  0.496  $\pm$0.051  \\
Gaia16alq  &  0.744  $\pm$0.055  & 0.768  $\pm$0.098 &  0.781  $\pm$0.07  &  0.764  $\pm$0.074  &  0.796  $\pm$0.08  \\
SN~2016asf  &  0.597  $\pm$0.149  & ... &  0.64  $\pm$0.055  &  0.618  $\pm$0.102  &  0.602  $\pm$0.064  \\
SN~2016bln  &  0.789  $\pm$0.097  & 0.802  $\pm$0.094 &  0.822  $\pm$0.071  &  0.804  $\pm$0.087  &  0.771  $\pm$0.088  \\
SN~2016coj  &  0.401  $\pm$0.053  & ... &  0.398  $\pm$0.035  &  0.4  $\pm$0.044  &  0.416  $\pm$0.047  \\
SN~2016eoa  &  0.482  $\pm$0.103  & ... &  0.507  $\pm$0.05  &  0.494  $\pm$0.076  &  0.497  $\pm$0.127  \\
SN~2016ffh  &  0.573  $\pm$0.078  & ... &  0.527  $\pm$0.125  &  0.55  $\pm$0.102  &  0.668  $\pm$0.107  \\
SN~2016gcl  &  0.689  $\pm$0.164  & 0.692  $\pm$0.102 &  0.719  $\pm$0.098  &  0.7  $\pm$0.121  &  0.642  $\pm$0.123  \\
SN~2016ixb  &  0.483  $\pm$0.064  & ... &  0.525  $\pm$0.098  &  0.504  $\pm$0.081  &  0.443  $\pm$0.088  \\
SN~2017cts  &  0.539  $\pm$0.063  & 0.558  $\pm$0.081 &  0.553  $\pm$0.099  &  0.55  $\pm$0.081  &  0.573  $\pm$0.089  \\
SN~2017erp  &  0.975  $\pm$0.083  & ... &  1.074  $\pm$0.12  &  1.024  $\pm$0.102  &  0.836  $\pm$0.087  \\
SN~2017fgc  &  0.692  $\pm$0.047  & ... &  0.701  $\pm$0.061  &  0.696  $\pm$0.054  &  0.695  $\pm$0.077  \\
SN~2017fms  &  0.36  $\pm$0.029  & ... &  0.385  $\pm$0.034  &  0.372  $\pm$0.032  &  0.367  $\pm$0.046  \\
SN~2017hjy  &  0.688  $\pm$0.057  & 0.69  $\pm$0.065 &  0.7  $\pm$0.069  &  0.693  $\pm$0.064  &  0.661  $\pm$0.081  \\
SN~2017igf  &  0.42  $\pm$0.051  & 0.409  $\pm$0.043 &  0.447  $\pm$0.04  &  0.425  $\pm$0.045  &  0.418  $\pm$0.057  \\
SN~2018oh  &  0.598  $\pm$0.059  & 0.614  $\pm$0.081 &  0.566  $\pm$0.073  &  0.593  $\pm$0.071  &  0.588  $\pm$0.084  \\
\hline
\end{tabular}
\caption{Initial $^{56}$Ni masses from the different methods used in this paper. $M_{ \rm Ni}^{\rm Arnett}$ is based on Eq.~\ref{eq:3}, collected from \citet{konyvestoth20}. $M_{ \rm Ni}^{\rm tail}$ and $M_{ \rm Ni}^{\rm t15}$ refers to the tail-mass and $t_{15}$-mass, respectively, as shown in Section~\ref{sec:results}. $M_{ \rm Ni}^{\rm avg}$ is the average of the previous three columns, while $M_{ \rm Ni}^{\rm KK}$ refers to the masses inferred from Equation~\ref{eq:6}. }
\label{tab:2}
\end{table*}

\begin{figure}
    \centering
    \includegraphics[scale=0.5]{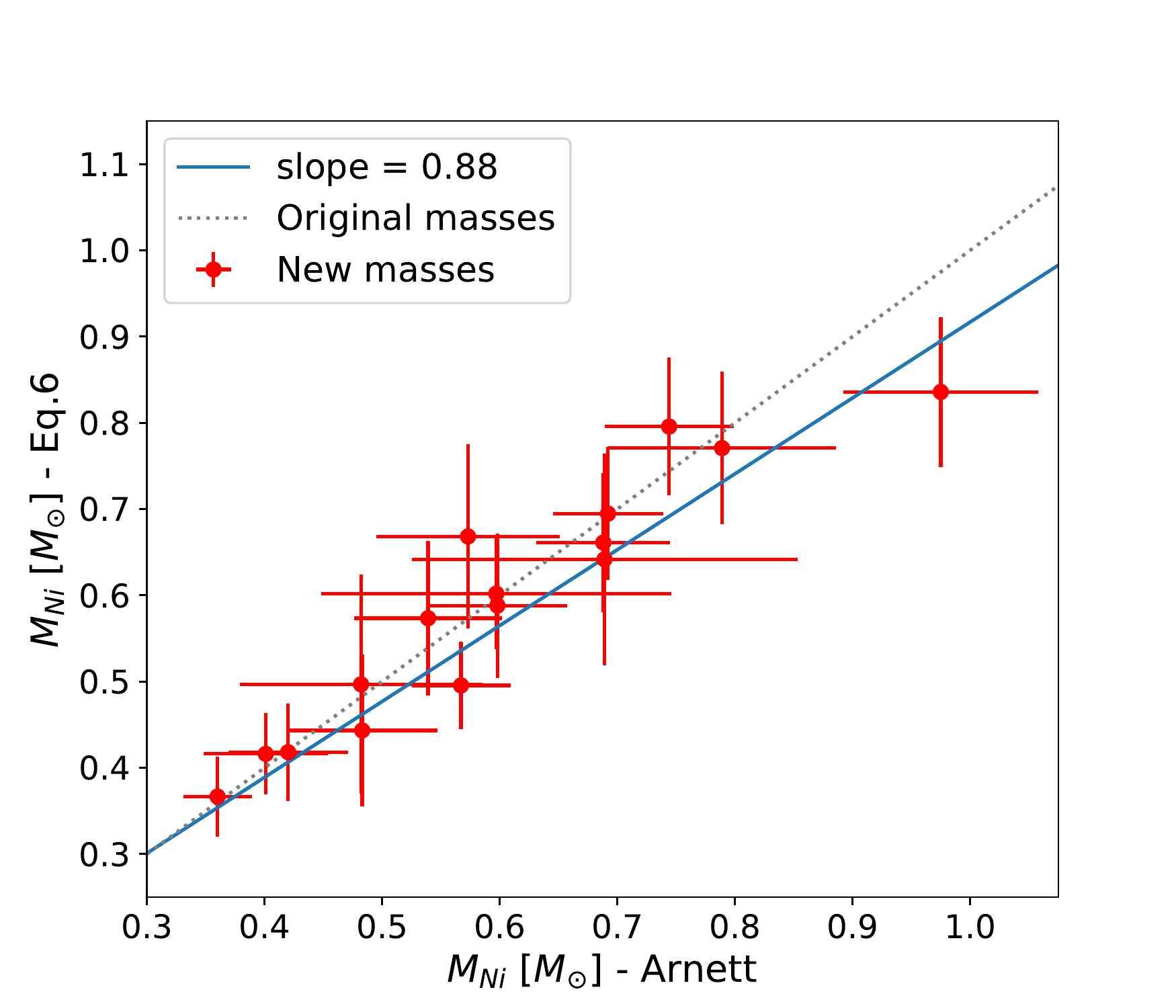}
    \caption{Comparison of the new nickel masses derived from Equation~\ref{eq:6} to the original ones from Arnett's rule in Table~\ref{tab:1}. {\bf The dotted line shows the 1:1 relation, while the solid line is the best-fit linear relation between the two datasets.}}
    \label{fig:2}
\end{figure}

Figure~\ref{fig:2} shows the comparison between the masses derived from Arnett's rule, and the ones estimated using Equation~\ref{eq:6}.

These new 
nickel masses scatter somewhat around the ones derived by 
\citet{konyvestoth20}
from Arnett's rule, but the differences are within the uncertainties, so the two datasets are generally consistent.
The SN that deviates the most from the previous estimates is SN~2017erp, which may have an overestimated $M_{ \rm Ni}$ due to its peculiar reddening \citep[see also][]{konyvestoth20}. Its early red color was also reported by \citet{li21}, and even though it was classified as a normal-velocity (NV) supernova, they also noted that it had the highest early-phase velocity within the group. Thus, if the nickel mass of SN~2017erp was previously overestimated, the new $M_{ \rm Ni}$ based on equation \ref{eq:6} might be a better estimate, since it is lower than the previous one from Arnett's rule.

The best-fit linear relation between $M_{\rm Ni}^{\rm Arnett}$ and $M_{\rm Ni}^{\rm KK}$, plotted with a solid blue line in Figure~\ref{fig:2}, has a slope of $0.880 \pm 0.068$, but if we omit SN~2017erp from the sample the slope becomes $0.967 \pm 0.075$, which is very close to the 1:1 relation (shown by the dotted line).

\begin{figure}[ht]
    \centering
    \includegraphics[width=8cm]{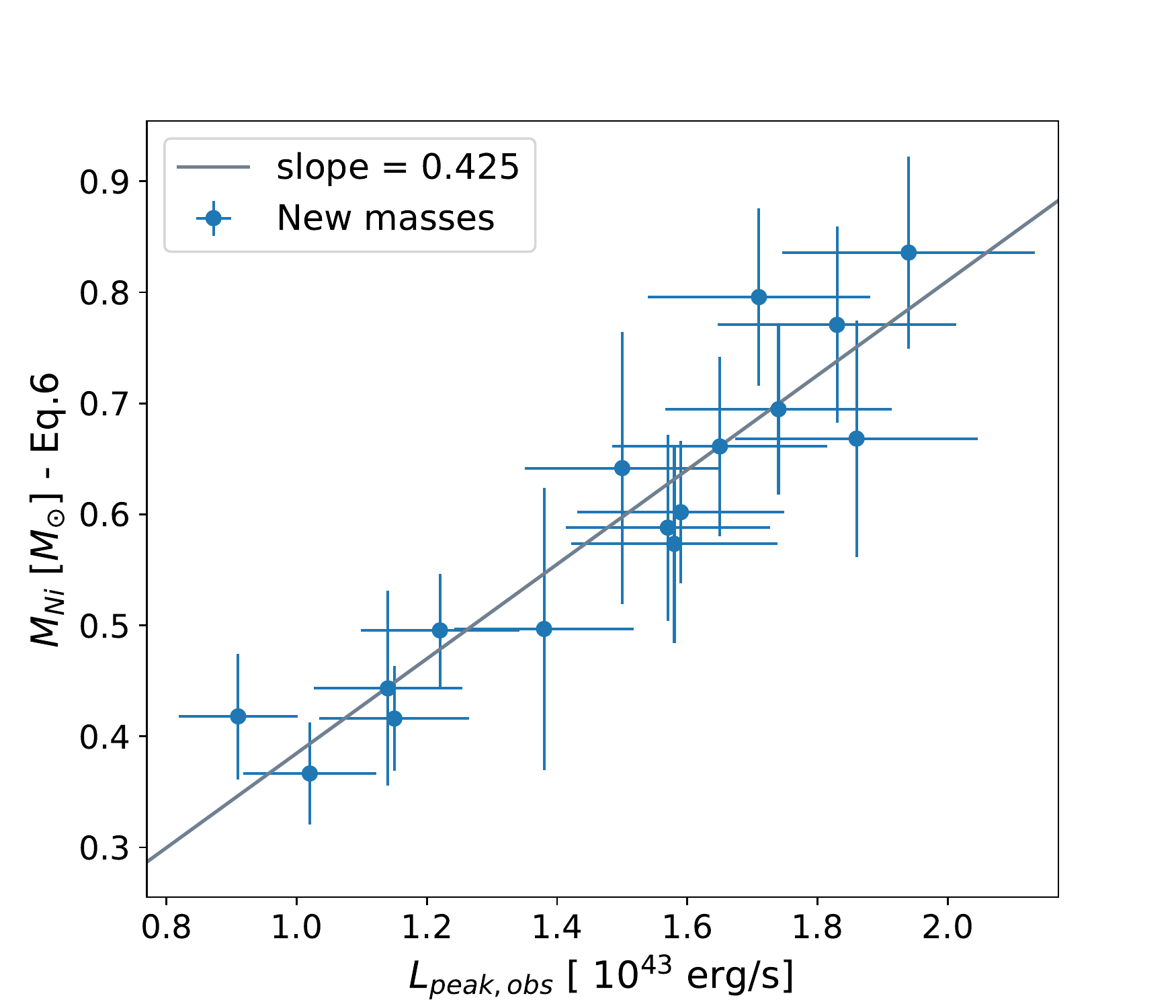}
    \caption{The new $^{56}$Ni masses against the observed peak luminosities.}
    \label{fig:3}
\end{figure}

In Figure~\ref{fig:3} we show the linear relationship between $L_{\rm peak, obs}$ and the new nickel masses {\bf ($M_{\rm Ni}^{\rm KK}$)}. It also demonstrates the diversity of the peak luminosities and the corresponding nickel masses, similar to \citet{scalzo14a, scalzo14b, scalzo19} and \citet{konyvestoth20}. A linear fit to these data resulted in a relation of 
\begin{equation}
    M_{\rm Ni}^{\rm KK} ~=~ 0.425 (\pm 0.038) \cdot L_{\rm peak,obs} ~-~ 0.041 (\pm 0.053),
    \label{eq:7}
\end{equation}
which can be used to estimate the initial nickel masses directly from the measured peak luminosities.

The validity of the new nickel masses estimated above is probed by using the published data on SN~2011fe. SN~2011fe was one of the most thoroughly studied SNe~Ia in the last decade, because it was a nearby, very bright event discovered only a few hours after explosion. There are numerous nickel mass estimates for SN~2011fe in the literature based on different methods. For example, \citet{pereira13} estimated $M_{\rm Ni}$ from spectrophotometric observations, similar to \citet{mazzali15}, who used optical and NIR spectra to determine the initial nickel and iron masses of the ejecta of SN~2011fe. \citet{scalzo14a} determined $M_{\rm Ni}$ from the peak bolometric luminosity, similar to our approach, while \citet{childress15} used the flux of the [CoIII] $\lambda$5893 nebular emission feature. More recently, \citet{dhawan16} related the phase of the secondary maximum of the near-infrared (NIR) light curves to the bolometric peak luminosity, from which they applied Arnett's rule and delayed-detonation models to determine the initial $M_{\rm Ni}$. As noted earlier, \citet{konyvestoth20} also gave an estimate for $M_{\rm Ni}$ by fitting the Arnett-model to the whole bolometric light curve.  All of these values are collected in Table~\ref{tab:3}, and their mean is $M_{ \rm Ni} = 0.50 \pm 0.08 M_{\rm \odot}$. This is within a $1 \sigma$ agreement with the new value of $M_{\rm Ni}^{KK} =  0.496 \pm 0.051 M_{\rm \odot}$ given above in Table~\ref{tab:2}.

\begin{table}
\centering
\begin{tabular}{cc} 
\hline
  $M_{ \rm Ni}$  &  Source \\
  $[M_{ \rm \odot}]$  & \\
\hline
0.53 $\pm$ 0.11   &    \citet{pereira13} \\
0.42 $\pm$ 0.08   &    \citet{scalzo14a} \\
0.47 $\pm$ 0.05   &    \citet{mazzali15} \\
0.500 $\pm$ 0.026 &    \citet{childress15} \\
0.52 $\pm$ 0.15   &   \citet{dhawan16} \\
0.567 $\pm$ 0.054 &    \citet{konyvestoth20} \\
\hline
0.50 $\pm$ 0.08   &    mean \\
0.496 $\pm$ 0.05 &    this work \\
\hline
\end{tabular}
\caption{Estimates for the initial $^{56}$Ni mass of SN~2011fe from different methods.}
\label{tab:3}
\end{table}

\section{A closer look at the Beta parameter}\label{sec:results}

The new formula by \citet{khatami19} {\bf(Equation~\ref{eq:6})} also introduces the $\beta$ parameter, which is connected with the spatial distribution of heating, recombination effects, and opacity. They found that  the different distributions of heating only changes the value of $\beta$, thus, {\bf Equation \ref{eq:6}} still holds true. This means that the same formula, with different $\beta$ values and heating functions, can be used to describe the peak luminosity of a wide variety of objects. 

{\bf If we have independent measurements for $M_{\rm Ni}$ and $L_{\rm peak}$ then we can apply Equation~\ref{eq:6} to infer $\beta$ for each object.
The original value {\bf of $\beta$}, given by \citet{khatami19} for Type Ia SNe as $\beta \sim 1.6$, is based on numerical simulations of SN Ia explosions.}

In this section, we use two independent methods to determine the nickel masses, and consequently, the values of $\beta$ for our sample SNe.

\subsection{Tail luminosity method}

It is well-known \citep{valenti08, scalzo14a, afsariarchi21} that
the late-phase light curve (at $t>60$ days after explosion) for a SN ejecta powered by the Ni-Co-Fe radioactive decay can be expressed as:
\begin{equation}
    L = L_{\gamma}(1 - e^{-t_\gamma^2/t^2}) + L_{\rm pos,KE}, 
    \label{eq:8}
\end{equation}
where $t_\gamma$ is, again, the timescale for gamma-ray leakage, while $L_\gamma$ is the luminosity 
released in the form of gamma-rays:
\begin{equation}
    L_{\gamma} = \frac{M_{Ni}}{\rm M_\odot} \left( C_{Ni} e^{-\frac{t}{t_{Ni}}} + 0.968 \cdot C_{Co} e^{-\frac{t}{t_{Co}}} \right),
    \label{eq:9}
\end{equation}

$L_{\rm pos,KE}$ gives the luminosity due to the thermalization of the kinetic energy of positrons released during the Co-decay:
\begin{equation}
    L_{\rm pos,KE} = \frac{M_{Ni}}{\rm M_\odot} \left[ 0.032 \cdot C_{Co} \left( e^{-\frac{t}{t_{Co}}} - e^{-\frac{t}{t_{Ni}}} \right)  \right].
    \label{eq:10}
\end{equation}

Comparing these formulae with Equation~\ref{eq:2}, we set
$t_{Ni} = 8.8$ days, $t_{Co} = 111.3$ days, $C_{Ni} = \varepsilon_{\rm Ni} - \varepsilon_{\rm Co} = 6.45 \cdot 10^{43}$ erg~s$^{-1}$ and $C_{Co} = \varepsilon_{\rm Co} = 1.45 \cdot 10^{43}$ erg~s$^{-1}$ \citep[see also][]{bw17}. 
$t_\gamma$ was given in Table~3 by \citet{konyvestoth20}, so the only free parameter in fitting the light curve tail is $M_{Ni}$. A caveat of this method is that it provides appropriate nickel masses only for data beyond $t \sim 60$ days, thus, we could obtain nickel masses only for those objects that had the late part of their light curve covered by observations. We list the results of fitting Equation~\ref{eq:8} to our data 
in Table~\ref{tab:2} as $M_{\rm Ni}^{\rm tail}$.

\subsection{The $t_{15}$ method}

\citet{sukhbold19} found that the bolometric luminosity is equal to $L_\gamma$ at $t_{15} = t_{peak} + 15$ days, similar to Arnett's rule. Using this, it is possible to determine the nickel mass by measuring the bolometric luminosity at $t_{15}$:
\begin{equation}
     L_{bol}(t_{15}) \approx \frac{M_{Ni}}{\rm M_\odot} \left( C_{Ni} e^{-\frac{t_{15}}{t_{Ni}}} + C_{Co} e^{-\frac{t_{15}}{t_{Co}}} \right) (1 - e^{-t_\gamma^2/t_{15}^2})
    \label{eq:11}
\end{equation}

$L_{bol}(t_{15})$ for each SN was determined by interpolating the bolometric light curves to $t_{15}$. The nickel masses found this way can be seen in Table~\ref{tab:2} as $M_{\rm Ni}^{\rm t15}$. 

\begin{figure}[ht]
    \centering
    \includegraphics[scale=0.5]{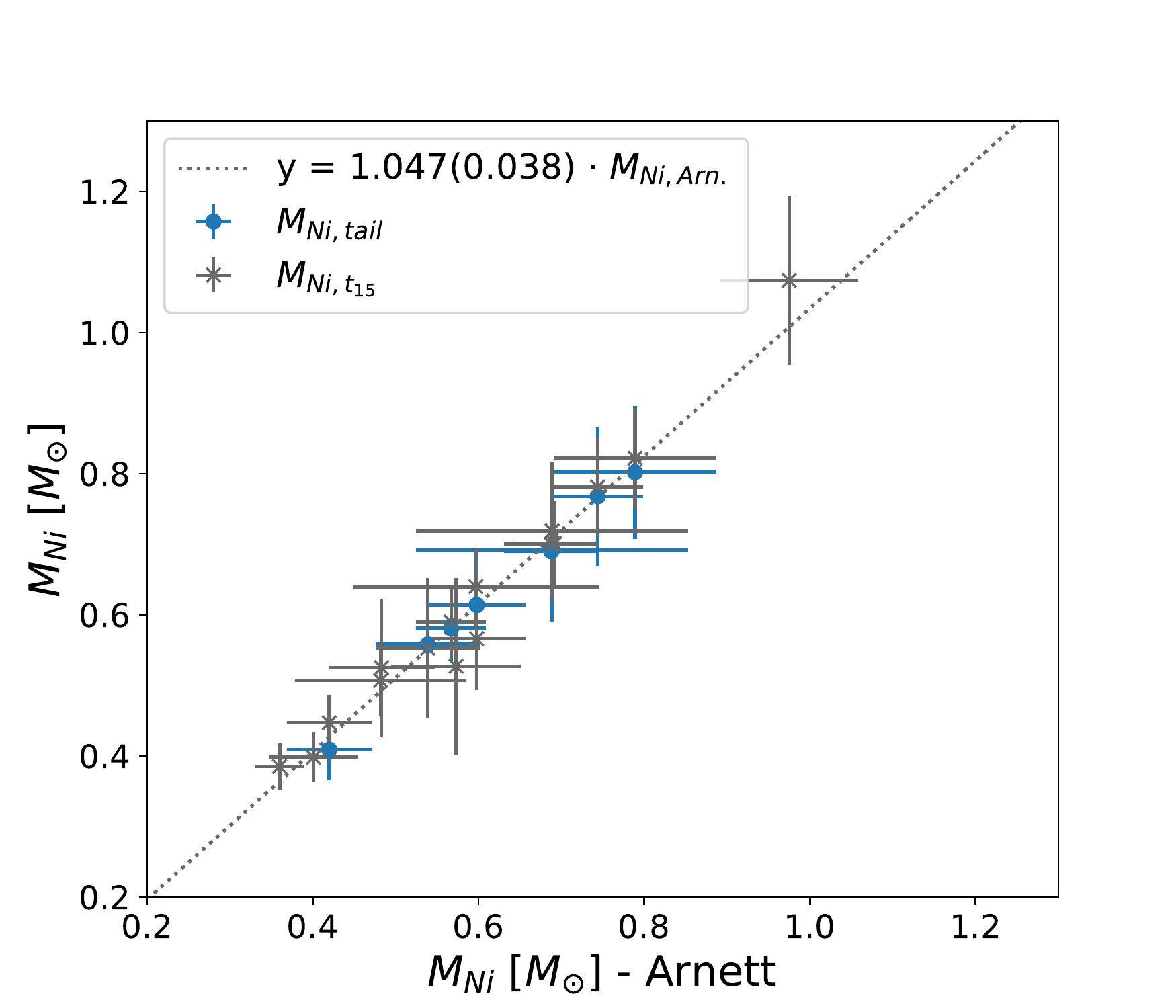}
    \caption{Values of $M_{Ni}$ from the tail luminosity and the $t_{15}$ methods compared to the masses from Arnett's rule.}
    \label{fig:4}
\end{figure}

It is seen from Table~\ref{tab:2} that the nickel masses inferred from both the tail luminosity method and the $t_{15}$ method are very similar to those obtained from Arnett's rule. Their consistency is further illustrated in Figure~\ref{fig:4}, where $M_{\rm Ni}^{\rm tail}$ and $M_{\rm Ni}^{\rm t15}$ are plotted against $M_{\rm Ni}^{\rm Arnett}$. All data points are closer to the 1:1 relation (shown as a dotted line) within their uncertainties.  The slope of the best-fit linear relationship, $1.047 \pm 0.038$, is consistent with the identity relation (see Fig.~\ref{fig:4}). 

\begin{figure}[]
    \centering
    \includegraphics[scale=0.5]{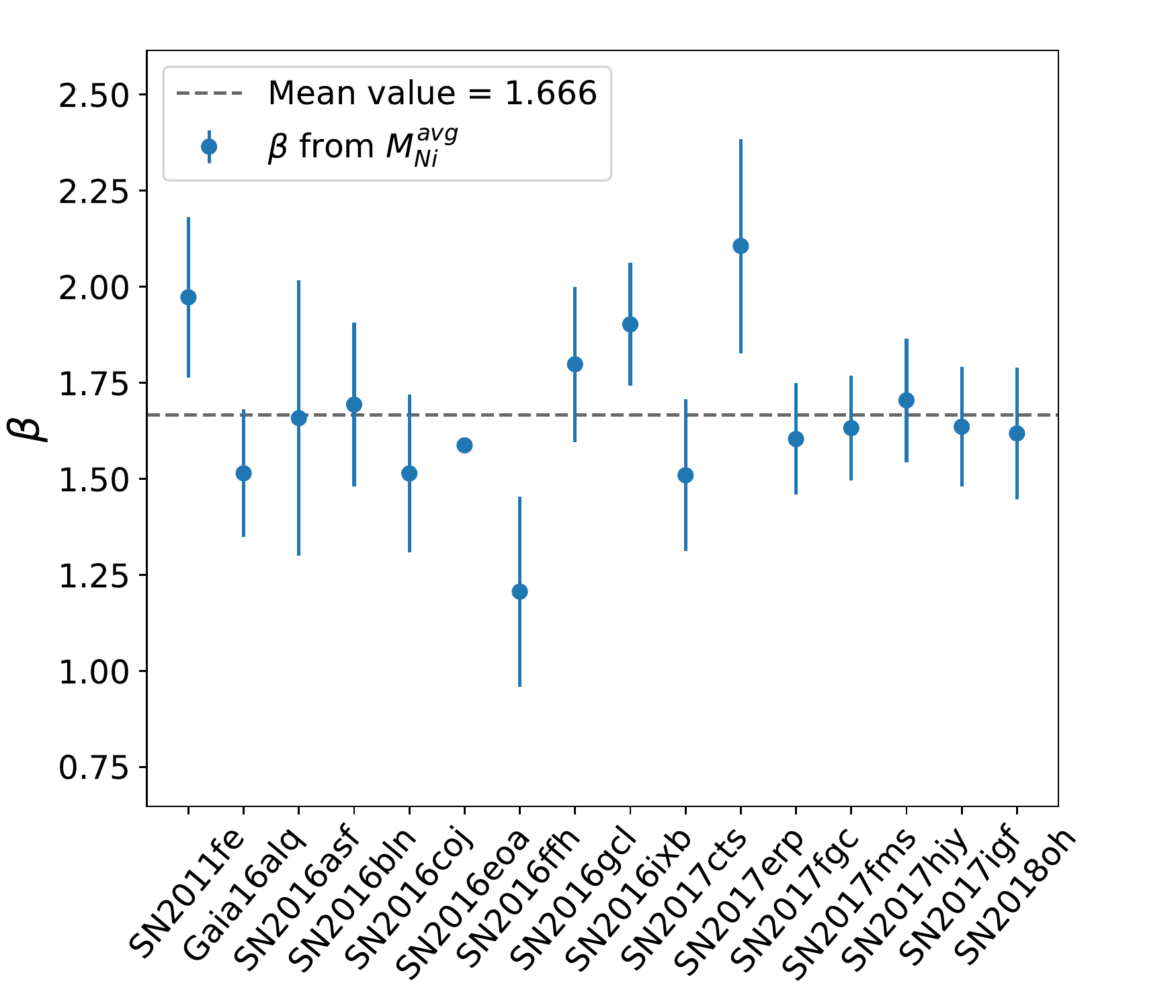}
    \caption{The new $\beta$ values inferred for the sample SNe.}
    \label{fig:5}
\end{figure}

\begin{figure}[]
    \centering
    \includegraphics[scale=0.5]{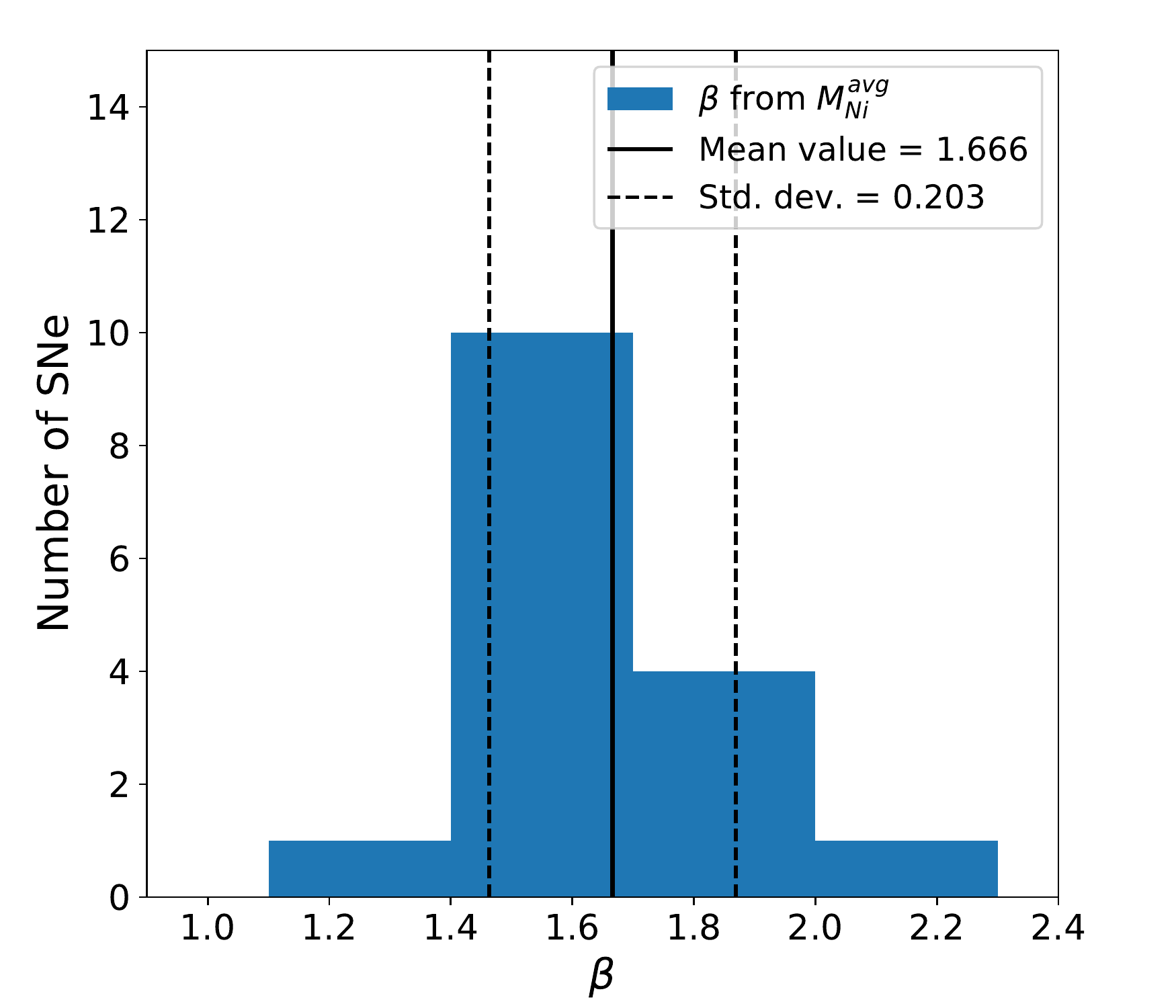}
    \caption{Distribution of $\beta$ values. The vertical thick and dotted lines show the mean value and the standard deviation of the sample, respectively.}
    \label{fig:6}
\end{figure}

Since all nickel masses found from different methods agree  within their uncertainties, we use their mean value (shown as $M_{\rm Ni}^{\rm avg}$ in Table~\ref{tab:2}) for measuring the $\beta$ parameter. 
We find that $\beta$ varies between 1.2 to 2.1 (Table~\ref{tab:4}), while the mean value is 1.666 $\pm$ 0.188, which is reasonably close (within $1 \sigma$) to 1.6 proposed by \citet{khatami19}. The inferred $\beta$ values are plotted in Figure~\ref{fig:5} for each SN, while their distribution is shown in Figure~\ref{fig:6}.

\begin{table}
\centering
\begin{tabular}{cc}
\hline
Object   &	$\beta$   \\
\hline
SN~2011fe  &  1.973  $\pm$  0.209 \\
Gaia16alq  &  1.515  $\pm$  0.166 \\
SN~2016asf  &  1.658  $\pm$  0.359 \\
SN~2016bln  &  1.693  $\pm$  0.214 \\
SN~2016coj  &  1.514  $\pm$  0.206 \\
SN~2016eoa  &  1.587  $\pm$  0.004 \\
SN~2016ffh  &  1.207  $\pm$  0.247 \\
SN~2016gcl  &  1.798  $\pm$  0.202 \\
SN~2016ixb  &  1.902  $\pm$  0.160 \\
SN~2017cts  &  1.509  $\pm$  0.198 \\
SN~2017erp  &  2.106  $\pm$  0.279 \\
SN~2017fgc  &  1.604  $\pm$  0.145 \\
SN~2017fms  &  1.632  $\pm$  0.136 \\
SN~2017hjy  &  1.705  $\pm$  0.161 \\
SN~2017igf  &  1.636  $\pm$  0.156 \\
SN~2018oh  &  1.618  $\pm$  0.172 \\
\hline
\end{tabular}
\caption{New individual $\beta$ values for the sample SNe}
\label{tab:4}
\end{table}

Comparing the results shown in Figures~\ref{fig:2} and \ref{fig:3} with that of a similar analysis published recently by \citet{afsariarchi21} for stripped-envelope supernovae (SESNe, see their Figure~4), one can see the essential difference between SESNe and Type Ia SNe: Arnett'rule significantly overestimates $M_{\rm Ni}$ with respect to $M_{\rm Ni}^{\rm tail}$ and $M_{\rm Ni}^{\rm KK}$ for SESNe, while all Ni masses agree very well with each other in Type Ia SNe. \citet{khatami19} also noted that, unlike in other SNe,  Arnett's rule seems to work quite well for Type Ia SNe due to their moderate (not too central, not too shallow) $^{56}$Ni distribution. Our empirical results presented here confirm this statement.

\section{Conclusions}\label{sec:conclusion}

We use previously published data \citep{konyvestoth20} to give estimates for the initial masses of radioactive nickel in 16 Type Ia supernovae using a new formula published by \citet{khatami19}, which relies on the relationship between the peak luminosity and peak time without assuming self-similar energy distribution within the ejecta. We compare our results with previous nickel mass estimates for SN~2011fe from the literature (see Table~\ref{tab:3}), and find very good agreement. Our new nickel masses are in a $1 \sigma$ agreement with those derived by others.

Previous estimates for the initial nickel mass in SNe Ia were mostly carried out by using Arnett's rule. Our results (Figure~\ref{fig:3}) show that the new formula by \citet{khatami19} gives consistent nickel masses with those estimated from a radiation-diffusion Arnett-model, while also taking into account the spatial distribution of heating that can be different in each case. Similar to \citet{scalzo19}, and \citet{konyvestoth20}, we find that the $^{56}$Ni masses show diversity, suggesting that the ejecta masses are also inhomogeneous. 

Finally, we give an approximate estimate for the $\beta$ parameter of each studied SN (Figures~\ref{fig:5} and \ref{fig:6}), and find good agreement with the mean value (1.6) given by \citet{khatami19} from SN Ia simulations. 

\acknowledgments{
This work is part of the project ``Transient Astrophysical Objects" GINOP 2.3.2-15-2016-00033 of the National Research, Development and Innovation Office (NKFIH), Hungary, funded by the European Union, 
{\bf and it was also supported by the NKFIH/OTKA FK-134432 grant.}
We thank the anonymous referee for the useful comments and suggestions that helped improving our paper.}

\software{Numpy \citep{numpy}, Matplotlib \citep{matplotlib}, Astropy \citep{astropy13} }

\end{document}